\input harvmac.tex
\input epsf

\ifx\epsfbox\UnDeFiNeD\message{(NO epsf.tex, FIGURES WILL BE IGNORED)}
\def\figin#1{\vskip2in}
\else\message{(FIGURES WILL BE INCLUDED)}\def\figin#1{#1}\fi
\def\tfig#1{{\xdef#1{Fig.\thinspace\the\figno}}
Fig.\thinspace\the\figno \global\advance\figno by1}
%

\def\IB{\relax\hbox{$\inbar\kern-.3em{\rm B}$}}
\def\IC{\relax\hbox{$\inbar\kern-.3em{\rm C}$}}
\def\ID{\relax\hbox{$\inbar\kern-.3em{\rm D}$}}
\def\IE{\relax\hbox{$\inbar\kern-.3em{\rm E}$}}
\def\IF{\relax\hbox{$\inbar\kern-.3em{\rm F}$}}
\def\IG{\relax\hbox{$\inbar\kern-.3em{\rm G}$}}
\def\IGa{\relax\hbox{${\rm I}\kern-.18em\Gamma$}}
\def\IH{\relax{\rm I\kern-.18em H}}
\def\IK{\relax{\rm I\kern-.18em K}}
\def\IL{\relax{\rm I\kern-.18em L}}
\def\IP{\relax{\rm I\kern-.18em P}}
\def\IR{\relax{\rm I\kern-.18em R}}
\def\IZ{\relax\ifmmode\mathchoice
{\hbox{\cmss Z\kern-.4em Z}}{\hbox{\cmss Z\kern-.4em Z}}
{\lower.9pt\hbox{\cmsss Z\kern-.4em Z}}
{\lower1.2pt\hbox{\cmsss Z\kern-.4em Z}}\else{\cmss Z\kern-.4emZ}\fi}

\def\CN {{\cal N}}

\def\CP {{\cal P}}
\def\CQ {{\cal Q}}

\def\CV {{\cal V}}


\def\p{\partial}



\def\Tr{{\rm Tr}}

\def\p{\partial}

\def\np{\nabla_{\partial}}

\def\ra{\rightarrow}

\def\inbar{\,\vrule height1.5ex width.4pt depth0pt}
\font\cmss=cmss10 \font\cmsss=cmss10 at 7pt

\def\a{\alpha}

\def\g{\gamma}
\def\d{\delta}

\def\m\mu
\def\n{\nu}

\def\p{\partial}
\def\R{\relax{\rm I\kern-.18em R}}
\font\cmss=cmss10 \font\cmsss=cmss10 at 7pt
\def\Z{\relax\ifmmode\mathchoice
{\hbox{\cmss Z\kern-.4em Z}}{\hbox{\cmss Z\kern-.4em Z}}
{\lower.9pt\hbox{\cmsss Z\kern-.4em Z}}
{\lower1.2pt\hbox{\cmsss Z\kern-.4em Z}}\else{\cmss Z\kern-.4em Z}\fi}
\def\pl{{\it  Phys. Lett.}}

\def\mpl{{\it Mod. Phys.   Lett.}}
\def\np{{\it Nucl. Phys. }}
\def\pr{{\it Phys.Rev. }}
\def\cmp{{\it Comm. Math. Phys. }}

\def\t{\theta}

\def\l{\lambda}

\def\ra{\rightarrow}

 %
  %


\lref\EK{ T. Eguchi and H. Kawai, Phys. Rev. Lett. 48 (1982) 1063.       }
\lref\BHN{ G. Bhanot, U. Heller and H. Neuberger, \pl 113B (1982) 47.  }
\lref\PARI{ G. Parisi, \pl 113B (1982) 463}
\lref\GK{ D. Gross and Y. Kitazawa, \np B206 (1982) 440.}
\lref\GA{ A. Gonsalez-Arroyo and M. Okawa, \pl B120 (1983) 174, 
\pr D27 (1983) 2397.}
\lref\DAVID{ F. David, \np B257 (1985) 45.}
\lref\KAZ{ V. A. Kazakov, \pl 150B (1985) 282.   }
\lref\WITTEN{   E. Witten, \np B460 (1995) 335, hep-th/9510135.  }
\lref\IDF{C. Itzykson and P. DiFranchesco, Ann. Inst. Henri Poincar\'e, 
vol.59, no. 2 (1993) 117.}
\lref\KSW{V. A. Kazakov, M. Staudacher and T. Wynter, \cmp 177 (1996) 451,
hep-th/9502132; \cmp 179  (1996) 235, hep-th/9506174; \np B440 (1995) 407,   
 hep-th/9601069.} 
\lref\BFSS{ Banks, W. Fischler, S. Shenker and L. Susskind, 
\pr D55 (1997) 5112, hep-th/9610043.}
\lref\IKKT{N. Ishibashi, H. Kawai, Y. Kitazawa and A. Tsuchiya, 
\np B498 (1997) 467, hep-th/9612115. }
\lref\IND{ S.~R.~Das, A.~Dhar, A.~Sentgupta and S.~R.~Wadia, 
\mpl A5 (1990) 41.}
\lref\NONC{ N.~Ishibashi, S.~Iso, H.~Kawai and Y.~Kitazawa, hep-th/9910004, 
hep-th/0001027. }
\lref\AMAK{J.~Ambjorn, Yu.~Makeenko, J.~Nishimura and R.~J.~Szabo, 
hep-th/9911041.}

\Title{}
{\vbox{
\centerline{  Field Theory as a  Matrix Model}
 \vskip2pt
}}
%
%
%
%
\centerline{Vladimir Kazakov \footnote{$ ^\bullet $}{{\tt
kazakov@physique.ens.fr}}}
\centerline{{\it $^1$  Laboratoire de Physique Th\'eorique de l'Ecole
Normale Sup\'erieure \footnote{$ ^\ast $}{
Unit\'e Mixte du
Centre National de la Recherche Scientifique 
et de l'Ecole Normale Sup\'erieure.}}}

\centerline{{\ \ \ \it  75231 Paris CEDEX, France}}

 \vskip 1cm
\baselineskip8pt{
 
\vskip .2in
 
\baselineskip8pt{
A new formulation of four dimensional quantum field theories, such as
 scalar field theory, is proposed as a large $n$ limit of a special
 $n\times n$ matrix model.  Our reduction scheme works beyond planar
 approximation and applies for QFT with finite number of fields. It
 uses quenched coordinates instead of quenched momenta of the old
 Eguchi-Kawai reduction known to yield correctly only the planar
 sector of quantum field theory. Fermions can be also included.  }

\bigskip
 \rightline{ LPTENS-00/10}

\Date{February, 2000}

\baselineskip=20pt plus 2pt minus 2pt

\newsec{Introduction }

To reproduce the physical space-space time out of some more
fundamental variables, rather than introduce it explicitly, has been
always a tempting idea in the quantum field theory. While the string
theory sets up a fruitful framework in which the space-time is
dynamically created out of the fluctuating coordinates of the strings,
potentially not less fruitful may be the attempts to encode the
space-time, together with the quantum fields themselves, into the
dynamics of large fluctuating matrices. A quantum field theory or a
string theory should be described in this case in terms of specific
matrix integrals containing finite amount of matrices of infinite
size. The earliest proposal of this kind belongs to T. Eguchi and
H. Kawai \EK , followed by a few important precisions and
modifications \BHN ,\PARI ,\GK , \GA .  In these works the QFT with
$N\times N$ matrix valued fields can be reduced in the large N
(planar, or 't Hooft) limit to finite dimensional matrix integrals in
the same limit.  

Another successful enterprise of this kind was a formulation and
solution of non-critical string theories (associated with the two
dimensional quantum gravity in the presence of some matter fields) in
terms of $U(N)$ invariant matrix integrals \DAVID , \KAZ .

More recently, one of the most fruitful ideas of this kind in the
superstring theory was the proposal of E. Witten \WITTEN\ generalizing
an old idea of Chan-Paton to describe the low energy physics of branes
by various reductions of super Yang-Mills theory. The diagonal
components of the effective SYM vector potentials play the role of
space-time coordinates of branes there.

The concept of the physical space-time built out of discrete degrees
of freedom has become especially inspiring due to fundamental
questions in quantum gravity, such as microscopic explanation of the
thermodynamics of black holes.

Our discussion in this paper will be confined to the matrix model
formulation of quantum field theory. To set up the problem let us
recall that the old Eguchi-Kawai (EK) reduction 
\foot{Using this abbreviation we remember of course about the 
important modification of the original, not quite working, reduction
scheme of Eguchi and Kawai related to the quenching of momenta \BHN\
and twisting \GA } reproduces correctly only the planar sector of a
matrix field theory, whereas the non-planar corrections ($1/N$
expansion) were never incorporated into this scheme, let alone a
nonperturbative formulation of a QFT with finite number of field
components (finite N) in terms of some matrix model of a matrix or
matrices of infinite size. The reason for this difficulty is mostly
due to the lack of reduced momenta running around nontrivial cycles of
non-planar graphs in the EK reduction scheme.

The aim of the present paper is to propose a new formulation of a {\it
finite} component scalar field theory (finite $N$) in terms of an
$n\times n$ matrix integral in the limit $n\ra\infty$. One can say
that it incorporates all orders of the $1/N$ expansion and is in
principle a possible non-perturbative definition of the original
scalar field theory.  For example, the usual scalar $\phi^4$ theory
can be formulated as a one matrix integral in external matrix
sources. We will also show how to incorporate fermions into this
scheme.  Unfortunately, we did not find so far any natural way to
formulate the four dimensional QCD in this way.

Our construction is in some sense T-dual to the old EK scheme: we use
the diagonal matrix sources of quenched coordinates instead quenched
momenta of the EK scheme. As a consequence of it the original scalar
field ``lives'' on the graphs dual to feynman graphs of our matrix
model. To control the parameters of these graphs (say, to make them
exactly $\phi^4$ graphs) we apply the methods worked out in \IDF\ and
\KSW\ for the so called model of dually weighted graphs
(DWG).

We will be able to generalize our method to fermions and to their
yukawa interactions with the scalars, but for the moment we don't know
a natural way to introduce  gauge symmetry into our approach. So to
formulate QCD is an interesting challenge in our framework.

\newsec{ Difficulty with $1/N$ corrections in the EK reduction scheme} 

Let us remind the essence of the old (and unsolved, to our knowledge)
problem of $1/N$ corrections to the reduced version of planar field
theory.  We will mostly discuss a matrix version of scalar field
theory in the euclidean four dimensional space described by the action
\eqn\SCALAR{  S= N\int d^4x \tr\left( (\p_\mu \phi)^2+V(\phi)\right)  }
where $\phi_{nm}(x)$ is an $N\times N$ hermitian matrix field and
$V(\phi)=\sum_{k\ge 2}{1\over k}t_k\phi^k$ is a scalar potential.

The EK reduction (in the most natural, Parisi formulation
\PARI ) goes as follows: take the following particular dependence of
$\phi$ on the coordinates:
\eqn\REDPAR{ \phi_{mn}(x)=e^{-i (p_m\cdot x)}\phi_{mn}e^{i (p_n\cdot x)}, 
\ \ \ \ {\it no \ \ summation \ \ over \ \ m,n}   }
The action then takes the form
\eqn\REDP{  S= \CV \tr\left( [p_\mu, \phi]^2+V(\phi)\right)  }
where $\CV$ is the 4D volume of the physical space,
$p_\mu=diag(p^{(1)}_\mu,\cdots,p^{(N)}_\mu)$ are D diagonal matrices
of quenched momenta scattered uniformly \GK\ in a large 4D ``momentum
box'' of a size $\Lambda^4$, where $\Lambda$ is UV cutoff, and the
original functional integral over scalar field is replaced by a single
matrix integral over $x$-independent $\phi_{mn}$. The planar sectors
(leading large N approximation) of the two models, the original matrix
scalar field theory in 4D and the reduced 0D one matrix model,
coincide. It is immediately clear from the double line representation
of planar graphs in the reduced theory (fig.1): each face of such a
graph is associated with a closed index loop and with the momentum
variable $p^{(i)}_\mu$carrying the same index $i$; each double line
propagator $D^{ii'}_{jj'}={1\over N}{\d_{ii'}\d_{jj'}\over
(p_i-p_j)^2}$ depends on the difference of the momenta of adjacent
faces. 

 %
\vskip 50pt
\hskip 20pt
\epsfbox{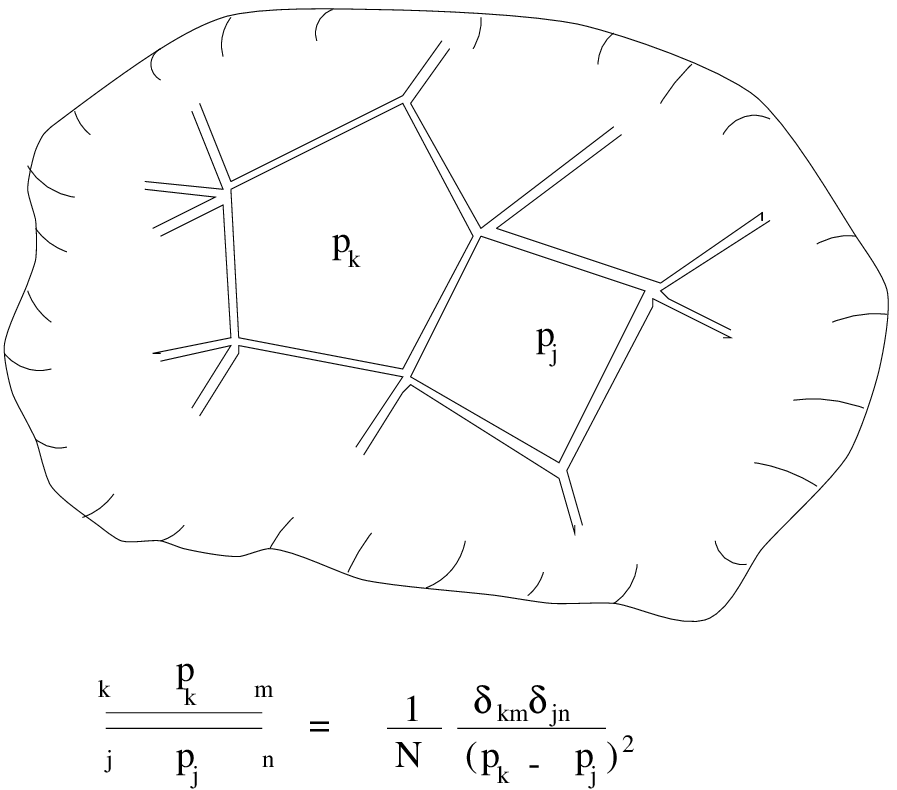}
\vskip 20pt
{\it  Fig.1:  A fragment of a planar diagram in the double line notation
and the propagator depending on the quenched momenta $p_k, \ p_j, \ k=1,
\cdots,N$ in Eguchi-Kawai reduction scheme.}

\bigskip

The planar part of the free energy of the reduced model
$F_{plan}={\it lim}_{N\ra\infty} {1\over N^2} \log Z_N$ ( $Z_N$ is the
partition function) can be schematically written in the following way:
\eqn\REDF{  F_{plan}=\CV  
\sum_G \prod_{v}t_{k_v}\left(\prod_{\tilde v}{1\over N}
\sum_{i_{\tilde v}}\right) \prod_{<{\tilde v}{\tilde v'}>} 
{1\over (p_{i_{\tilde v}}-p_{j_{\tilde v'}})^2} }
where $\sum_G$ goes over planar feynman graphs $G$ of the scalar field
theory, $v,\tilde v$ we label respectively original and dual vertices
of a graph $G$, $i_{\tilde v}1,\cdots,N$ is the index associated with
the dual vertex $\tilde v$ and $\prod_{<{\tilde v}{\tilde v'}>}$ goes
over the edges $<{\tilde v}{\tilde v'}>$ of the graph $\tilde G$ dual
to $G$.

In the large N limit only planar graphs survive in both models and the
sums over indices reproduce the integrals over 4D momenta: ${1\over
N}\sum_{i=1}^N\cdots\ra \int_{|p|<\Lambda} d^4p\cdots$.  Hence the
planar sectors of the original matrix scalar field theory and of its
reduced version are equivalent. This equivalence extends to any one
point $U(N)$ invariant physical quantities of the type $O_k=<{tr\over
N}\phi^k(x)>$ but is known to fail for the multi-point correlators
since $<{tr\over N}\phi^k(x){tr\over N}\phi^l(y) >=<{tr\over
N}\phi^k(x)><{\tr\over N}\phi^l(y)>$ plus $O(1/N^2)$ corrections,
different in two models. The EK reduction fails to describe correctly
the higher $1/N$ (non-planar) corrections to the original scalar field
theory.

The reason for this failure is well known: we cannot represent all
momenta running through the propagators as differences of momenta of
the adjacent loops (faces) on the graphs of a non-spherical topology. If
we did so (and it is precisely the case of the topological expansion
in the EK reduced model) the momenta running along topologically
nontrivial cycles of a non-planar feynman graph would be zero. To see
it one takes any nontrivial closed path on a dual graph (connecting
dual vertices or original faces) and calculates the total momentum
running through the propagators crossed by this path as
$(p_i-p_j)+(p_j-p_k)+\cdots+(p_l-p_i)\equiv 0$. For example, for the
torus topology we have two momenta of the original 4D theory missing
in the reduced version: they flow through two nontrivial cycles of the
torus.

The reason for the failure is simple but the remedy is not easy to
find, at least in case of the EK reduction involving reduced
momenta. 

In the next section we will show that the goal of construction of a
matrix model describing a finite $N$ (including the most frequent
$N=1$ case) scalar field theory can be achieved by introducing
quenched coordinates instead of quenched momenta.

\newsec{Matrix model formulation of finite $N$ scalar field theory}

Now we will propose a one matrix model in external matrix
fields which will be equivalent, at least perturbativly, graph by
graph of any topology, to the original 4D finite $N$ matrix scalar
field theory \SCALAR . We will show that the free energy and, with an
appropriate definition, the physical quantities of the field theory
\SCALAR\ at finite $N$ coincide with those of the matrix integral 
over a hermitian  $n\times n$ matrix $\Phi$ in the limit $n\ra\infty$:
\eqn\PART{  Z=e^{N^2 F}=\int d^{n^2}\Phi e^{-S}  }
with the action:
\eqn\REDX{S= N\Tr_{\CN}\left( [X_\mu,\Phi]^2+\ln(I_\CN- A \Phi)\right)  }
Here the matrix $\Phi$ lives in the $n=p\cdot q$ dimensional vector
space $\CN$ which is a direct product $\CN=\CP\times\CQ$ of vector
spaces of smaller dimensions $p$ and $q$, correspondingly.  $I_n$ is
the $n\times n$ unity matrix. Both dimensions $p,q$ go to infinity as
$n\ra \infty$; $X_\mu$ and $A$ are external (fixed) diagonal matrices
of the form
\eqn\XMAT{ X_\mu=\hat x_\mu \times I_\CQ, } 
where 
$\hat x_\mu=diag(x^{(1)}_\mu,\cdots, x^{(p)}_\mu)$,
\eqn\AMAT{  A=I_{\CP}\times \hat a} 
where $\hat a=diag(a_1,\cdots a_q)$.

So the matrices $\hat x_\mu$ on the one hand and the matrix $\hat a$
on the other hand live in orthogonal subspaces.

The matrices $\hat x_\mu$ with $\mu=1,2,3,4$ will play the role of
quenched coordinates: the points with coordinates $x^{(i)}_\mu$ should
be distributed uniformly in the physical space box of a size $L^4$,
which is the size of our system. The ultraviolet cutoff is defined as
$\Lambda\sim{ p^{1/4}\over L}$. Obviously the thermodynamic limit of
(infinite volume) corresponds to $p\ra\infty$ with $\Lambda$ fixed.

The matrix $\hat a$ will encode the information about the couplings of
scalar potential
\eqn\POTV{  V(\phi)=\sum_{k\ge 2}{1\over k} \ t_k \ \phi^k  }
in the form:
\eqn\SIMP{  t_k=-{p\over N} \ \tr_Q \hat a^k }
where the trace $\tr_\CQ$ goes only with respect to the vector space
$\CQ$. Here $t_2\equiv m^2$ corresponds to the mass squared.

The parametrization \SIMP\ of the couplings reminds the so called Miwa
variables widely used in the theory of $\tau$-functions of the
hierarchies of integrable differential equations. It is also used in
the representation of characters of the group $GL(N)$ through Schur
polynomials (see \KSW\ for the details). 

 Obviously the last formula can be in general true only in the limit
$q\ra\infty$. Note that $N$ is kept as a finite fixed parameter here.
 
>From \POTV\ and \SIMP\ the potential can be also written in the form
\eqn\POLV{ V(z)=-{p\over N}\sum_{j=1}^q \ln( 1-a_j z)  }
In the limit $q\ra\infty$ we can parameterize in principle any
potential (including a polynomial one) by such a sum of logarithmic
terms, but $a_i$'s need not be necessary real. They can be taken, say,
in complex conjugate pairs (the potential becomes even in this case).

For instance we can choose $a_i$ in such a way that in the limit
$q\ra\infty$ they will reproduce any polynomial potential \POTV :
\eqn\ATHET{  a_j=e^{i\theta_j}   }
where
\eqn\THETK{\t_j={2\pi\over q}j +
{2N\over  p q}\sum_{m\ge 1}{t_m\over m}\sin{2\pi j  m\over q} }
Indeed, for $k>0$ we have
$$\sum_{j=1}^q e^{i k\theta_j} 
\simeq_{p,q\ra\infty}\sum_{j=1}^q e^{{2i\pi \over q}j k}
\left(1+ik{2N\over p q}\sum_{m\ge 1}{t_m\over m}
\sin{2\pi j m\over q} \right) =-{N\over p}t_k.$$

The proof of the equivalence of the QFT \SCALAR\ and the matrix model
\REDX\ is very simple.  Let us consider any feynman graph of the
theory \REDX\ (dotted line on fig. 2) together with its dual graph
(solid line on fig. 2). The matrix structure of the theory prescribes
to use the double line notations for the propagators, so such graph
has a fixed topology with the genus defined by the Euler formula. Each
single line carries now a double index corresponding to the product of
spaces $\CP\times\CQ$.

 %
\vskip 50pt
\hskip 20pt
\epsfbox{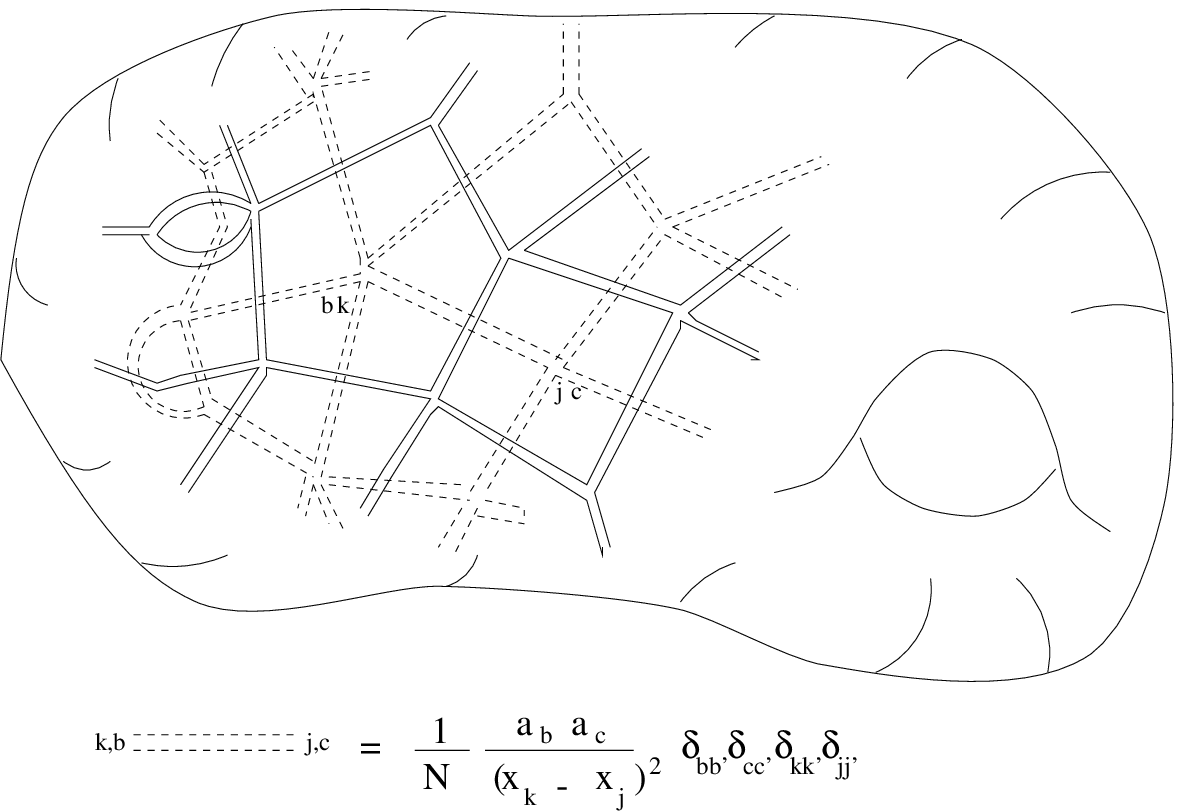}
\vskip 20pt
{\it Fig.2.  Original and dual graphs and the dual propagator
depending on the quenched coordinates $x_k,x_j$ in our one matrix
integral representation of scalar QFT. We show the pairs of indices
$b,k$ and $c,j$ belonging to $\CQ,\CP$ spaces, correspondingly,
running around original faces, and hence placed at dual vertices.}

\bigskip

The propagators of the original graph are given by:
$$ {1\over N}{a_b a_c \over (x_i-x_j)^2}
\d_{bb'}\d_{cc'}\d_{ii'}\d_{jj'}.$$ Here we attributed the $a_k$
factors to the propagators rather than to the vertices which can be
achieved by the change of the matrix variable: $\Phi \ra
A^{-1/2}\Phi A^{-1/2}$. It is easy to see that we obtain at each face
of the original graph a weight $\tr_\CQ \hat\a^k$ where k is the
order of this face (number of edges). It happens in the same way as in
the so called matrix model of dually weighted graphs (DWG) - a one
matrix model with the action $S=n\tr[\Phi^2+W(A \Phi)]$ studied in
\IND ,\IDF\ and solved in \KSW : apart from original couplings
coming from the potential $W(z)$ we also obtain in the DWG model the
dual couplings $\tilde t_k$ weighting the faces of different orders k
(or, which is the same, the vertices of the dual graph with the
coordination number k).

The factors ${1\over (x_i - x_j)^2}$ can be now attributed to the dual
propagators (crossing the original ones) and the indices $i,j$ are
running around the faces adjacent to the original propagators or, which
is the same, attributed to the corresponding dual vertices (see fig.2).

It is natural to formulate the result for the free energy in terms
of feynman expansion with respect to the dual graphs $\tilde G$. It
can be written in the following way:
\eqn\FREN{   F=\sum_{\tilde G}N^{2-2g}
\prod_{\tilde v}\big(t_{k_{\tilde v}}{1\over p}
\sum_{i_{\tilde v}=1}^p\big)\prod_{<\tilde v' \tilde v''>}
 \big(x_{i_{\tilde v'}} - x_{i_{\tilde v''}}\big)^{-2}        }
where $\sum_{\tilde G}$ goes over all dual graphs of the matrix model
\REDX , $g$ is the genus of a graph $\tilde G$, $\prod_{\tilde v}$ 
goes over all vertices $\tilde v$ of this graph with coordination
numbers $k$, $t_k=-{p\over N}\sum_{j=1}^q a_j^k$ are the couplings
attached to these vertices and $\prod_{<\tilde v' \tilde v''>}$ goes
with respect to all edges of $\tilde G$ connecting the vertices
$\tilde v$ and $\tilde v'$. At each dual vertex ${\tilde v}$ there is
a sum taken with respect to the index $i_{\tilde v}$ corresponding
to the subspace $\CP$.

Note also that due to the specific logarithmic form of the
interactions in \REDX\ $-\Tr\ln\big(I-A\Phi)= \sum_k {1\over
k}\Tr(A\Phi)^k$ all faces of dual graphs appear weighted with the
factor 1 (the factor ${1\over k}$ compensates the cyclic symmetry of
each dual face). So we see that \FREN\ is given by the sum over
connected graphs $\tilde G$ of all genera waited by $N^{2-2g}$, with
unrestricted face order and with the vertices waited by couplings
$t_{k_{\tilde v}}={ p\over N} \ \tr_\CQ \ \hat a^{k_{\tilde
v}}$. Hence these are precisely the original graphs of the scalar
matrix field theory \SCALAR\ with the appropriate 4D massless
propagators in the coordinate space. The mass $m$ is taken into
account by the presence of the coupling $t_2\equiv -m^2$ in the scalar
potential.

It is left to add that the summations over the indices $i_{\tilde
v}=1,\cdots,p$ can be substituted by the integrations in the large $p$
limit:
\eqn\SUMINT{   {1\over p}\sum_{i=1}^p\cdots \ra_{p\ra\infty} 
\int d^4x\cdots }

In this way we can reproduce the correspondence between the matrix
scalar 4D QFT \SCALAR\ and the zero dimensional matrix integral \REDX ,
graph by graph of any topology. Hence  they coincide, at least
in any order of the perturbation theory. 

Let us stress again that $N$ is a fixed parameter in our
construction. It does not even need to be integer, although integer
$N$ seems to be singled out by the fact that we can represent the
corresponding determinant as an integral over an $N$-vector of complex
$n\times n$ matrices $M_l, \ l=1,\cdots,N$:
$$
\exp[-N \Tr\log(I- A\Phi)]=
\int \prod_{l=1}^N d^{2N^2}\! M_l \ \exp[-\Tr M_l^+(I-A\Phi)M_l]
$$ 
 In particular, for $N=1$ \REDX\ is equivalent to the usual one
component scalar field theory with the action $S=\int d^4x \big
[ (\p_\mu\phi)^2+V(\phi)\big]$.

To calculate the one point correlators we use the following
correspondence between the averages in the scalar QFT and its matrix
model (MM) formulation (we take $N=1$):
\eqn\PONE{  <\int \ d^4x\phi^k(x)>_{QFT}=
{1\over \tr_\CN A^k}<\tr_\CN(A\Phi)^k>_{MM}     }
For the two point correlator we have the correspondence:
\eqn\PTWO{  <\phi(x)\phi(y))>_{QFT}={1\over (\tr_\CQ \hat a)^2}
<\left(\tr_\CN(A\Phi)\right)_{ii}\left(\tr_\CN(A\Phi)\right)_{jj}>_{MM}}
where the traces are taken only with respect to the subspace $\CQ$ and
the matrix indices $ii$ and $jj$ of the subspace $\CP$ are fixed and
chosen in such a way that $(x_i-x_j)^2\simeq (x-y)^2$.

\newsec{   Reduction in the presence of fermions and yukawa 
interactions }

Although we don't know any natural way to build a reduction of
Yang-Mills theory for finite $N$, the scalar field theory \SCALAR\ is
not the only interesting QFT which can be reduced to a matrix model in
this way.

Let us consider as an example of application of our method the
reduction of the QFT of massless Dirac fermions and massless bosons
with yukawa interaction in 4D with the action:
\eqn\YUKAWA{  S= \int d^4x \left\{ (\p_\mu \phi)^2
+\bar \psi\g_\mu\psi- {\l\over 3} \bar\psi\psi\phi \right\}  }
We stress that here $\phi(x)$ and $\psi(x)$ are usual bosonic and
dirac fields: $\phi$ has only one component and $\psi$ is a dirac
spinor. The reduced version of this model is given in terms of matrix
integral over 2 hermitian $n\times n$ matrices $\Phi$ and $\Psi$ in
the auxiliary linear space $\CN=\CQ\times \CP$
\eqn\PARTYU{  Z=e^F=\int d^{n^2}\Phi d^{n^2}\Psi \ e^{-S}  }
with the action
\eqn\REDYU{  S= \Tr_\CN\left( [X_\mu,\Phi]^2+[X_\mu,[X_\nu,\Psi]]^2  
-\tr_D\ln(I_\CN- \g_\mu\times[X_\mu,\Psi]A)
+\ln(I_\CN- A \Phi A\Psi)\right)     }
Here $\tr_D$ is taken with respect to the dirac indices of $\g_\mu$
matrix which is in the direct product with all other matrices in the
third term. The $A$ matrix is chosen in such a way that $p\sum_{j=1}^q
a_j^k=\l \delta_{k,3}$ for $k>0$ in the large $q$ limit. The last can
be achieved for example by the following choice of $\hat a$: $a_j=
\left({2\l \over p q}\right)^{1/3} e^{i\t_j}$ with $\t_j$ given by the
equation ${2\pi j\over q}=\t+{1\over 3}\sin3\t$. To prove it it is
enough to calculate the density of $\t$'s in the large $q$ limit:
$\rho(\t)={1\over q}{\p j\over \p\t}={1\over 2\pi}(1+{1\over3}\cos
3\t)$ which gives $\tr_\CQ \hat a^k = \int_{-\pi}^\pi
e^{ik\t}\rho(\t)=\l \delta_{k,3}$ for $k>0$.

All other definitions are the same as in the previous section.

To verify the perturbative equivalence of the QFT \YUKAWA\ and the
matrix integral \PARTYU -\REDYU\ we compare again the feynman graphs
of the former to the dual feynman graphs of the latter.  Due to the
choice of the matrix $\hat a$ dual graphs of \REDYU\ will contain only
triple interaction vertices and due to the $\log$ type potentials the
order of dual faces will be unrestricted (see fig. 3).

\vskip 50pt
\hskip 20pt
\epsfbox{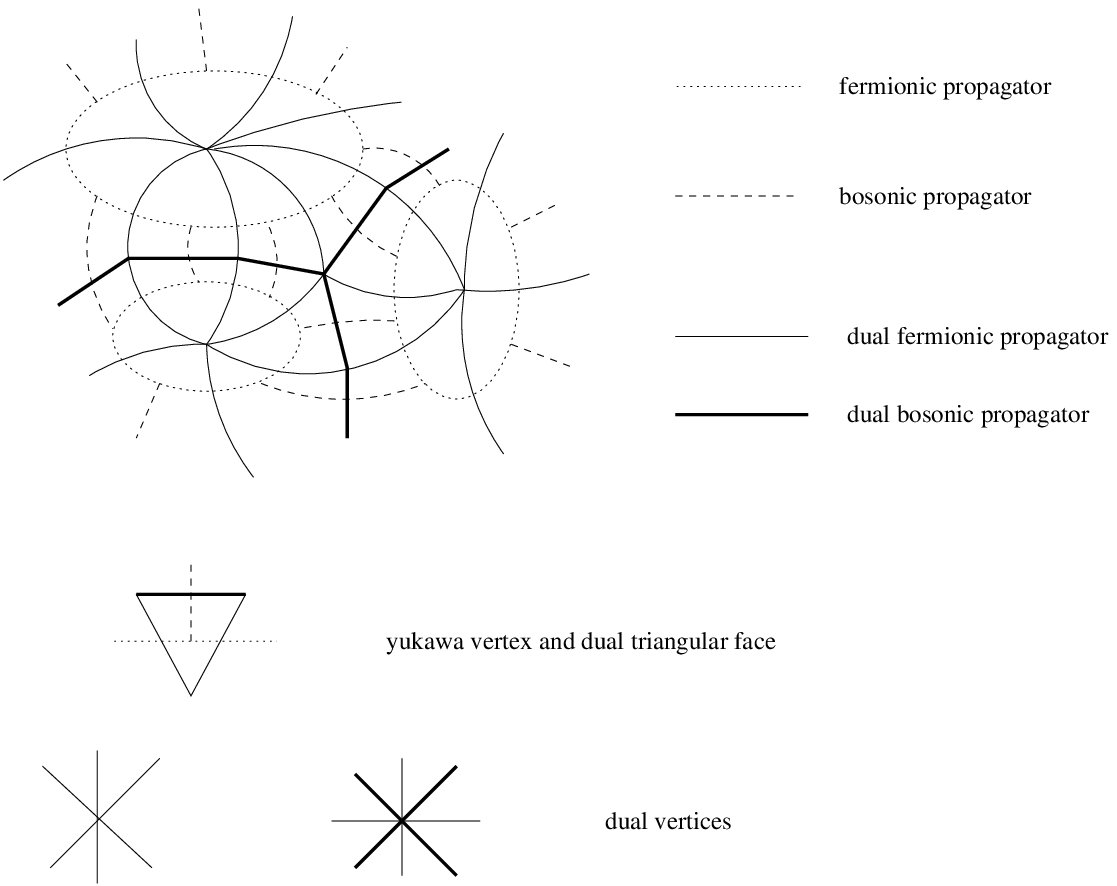}
\vskip 20pt
{\it Fig.3:  Original and dual graphs of the QFT of scalars and 
fermions with yukawa interactions (in single line notations). }

\bigskip

Note that feynman graphs of the QFT \YUKAWA\ have two types of faces:
fermionic loops built out of Dirac propagators and the loops with the
boundary built from interchanging dirac and bosonic propagators.  The
first type of loops will be generated by the third term in \REDYU\ and
the second kind - by the last term in \REDYU . The yukawa interaction
vertices correspond to the loops of dual graphs of the matrix model
\REDYU . Note the $(-)$ sign in front of the third term which gives
the fermionic statistics ($(-)$ sign for each fermionic loop). Each
fermionic loop of $k$ vertices is equipped, as it should be, by the
factor $\tr_D(\g_{\mu_1}\cdots\g_{\mu_k})$.  Finally, the massless 4D
fermionic propagator ${\g_\mu(x^\mu-y^\nu)\over |x-y|^4}$ will appear
as a function of difference of quenched coordinates
${\g_\mu(x_i^\mu-x_i^\nu)\over |x_i-x_j|^4}$.

Hence in the same limit $p,q\ra\infty$ as for the scalar QFT of the
previous section we reproduce graph by graph (with the weights
independent on the topology of graphs) the free energy of the QFT
\YUKAWA\ out of the matrix integral \REDYU .

\newsec{ Conclusions and comments }

We proposed a reduction of 4D quantum field theories with a finite
number of fields (finite $N$ in case of matrix fields) to a matrix
integral over infinite matrices. The physical four dimansional
coordinate space is encoded into the components of D=4 auxiliary
diagonal matrices of quenched coordinates. The reduction is different
from the old Eguchi-Kawai reduction with quenched momenta: it
reproduces correctly not only the planar approximation but also, at
least perturbatively graph by graph, all non planar corrections.  The
coordinate space is introduced in our reduction scheme in a way which
reminds the description of coordinates of D-branes in \WITTEN .  The
large $n$ limit that we use is different from the usual 'tHooft limit
and rather similar to the one adopted in \BFSS\ for the matrix model
of M-theory.

We showed that fermions can be also naturally introduced into this
reduction scheme.

A few comments are in order:

1. The QFT's in dimesions different from D=4 can be also reduced in
   this way but since the scalar propagator is different from
   $1/(x-y)^2$ the corresponding matrix representation of it looks
   ``nonlocal'' (i.e., different from $\Tr[X_\mu,\Phi]^2$). One can
   use the formula $\Tr[X_{\mu_1},[\cdots,[X_{\mu_k},\Phi]]\cdots]^2=
   \sum_{i,j}(x_i-x_j)^{2k}|\Phi_{ij}|^2$ in the action of the reduced
   matrix model to supply the dual graphs by any propagator
   $D(x_i-x_j)$.

2. Although our new matrix formulation of some QFT's hardly helps for
   solving them analytically it may provide a new numerical approach
   to their study: our matrix model does not deal with any 4D lattice
   and the approach to the thermodynamical limit might be much faster
   then in the conventional Monte Carlo algorithms based on the
   lattices. One could also envisage some real space renormalization
   schemes where the renormalization flow would be considered with
   respect to the size of matrices.

3. The authors of the papers \NONC , \AMAK\ propose a formulation of
   the non-comutative QFT using reduction to the matrix models with
   quenched momentum matrices $P_\mu$ obeying the Heisenberg
   commutation relations: $[P_\mu,P_\nu]=iB_{\mu\nu}$ where
   $B_{\mu\nu}$ is an antisymmetric $D\times D$ matrix of
   C-numbers. For our model, a natural non-comutative generalization
   would occur if we take the same commutation relations for the
   coordinate matrices $X_\mu$: $[X_\mu,X_\nu]=iC_{\mu\nu}$. Since the
   coordinate and momentum matrices are connected in \NONC\ by the
   relation $P_\mu=B_{\mu\nu}X_\nu$ it is natural to expect that we
   get the same non-comutative scalar field theory if we take the
   matrix $C=B^{-1}$. We haven't yet a proof of this statement.

Let us also mention the most obvious problems and questions concerning 
our matrix reduction of QFT's:

1. We did not manage to find a natural matrix reduction of QCD with 3
   colors. In principle we could try to do it in a particular gauge by
   systematically reproducing all feynman diagrams by the method we
   presented in this paper. But it would be much more instructive to
   find some general principle for such reduction arising explicitely
   from the gauge invariance. For the moment such a principle is
   missing.

2. It would be interesting to find a modification of our approach
   similar to the twisted EK model proposed in \GA . In \GA\ the
   momentum space appears as a classical vacuum of a large N lattice
   QFT with modified couplings. We can imagine a similar twisting of
   reduced matrix model with the classical vacuum solution generating
   the coordinate space.

3. A related but more ambitious question: can we build realistic
   models of fundamental interactions as some reduced matrix models
   where the physical space would emerge due to some symmetry breaking
   procedure? In other word, can our World be described by a Matrix
   Model? We know some of the attempts of this kind in the superstring
   theory and M-theory \BFSS , \IKKT . Our construction might be
   useful to approach this problem from a different direction.

\newsec{Acknowledgments}

%
\noindent  

The discussions with Hong Liu and Ivan Kostov were very useful for
this project.

I would like to thank the Physics Department of the Universidad
Catolica in Santiago (Chile) where this work was started, and
especially Jorge Alfaro for the kind hospitality.

 I am particularly grateful to the Weizmann Institute for Science in
Israel where most of this work was done during my stay there on the
Landau-Weizmann program and especially Shimon Levit and Adam Schwimmer
for their kind hospitality and fruitful interactions.

This research is supported in part by European TMR contract
ERBFMRXCT960012 and EC Contract FMRX-CT96-0012.


\listrefs
\bye